\input{aipcheck.tex}

\documentclass[final]{aipproc}

\layoutstyle{6x9}

\usepackage{epsfig}
\usepackage{amsfonts}

\newcommand{\lb}[1]{\label{#1}}

\newcommand{\bc}{\begin{center}}
\newcommand{\ec}{\end{center}}
\newcommand{\be}{\begin{equation}}
\newcommand{\ee}{\end{equation}}
\newcommand{\bea}{\begin{eqnarray}}
\newcommand{\eea}{\end{eqnarray}}
\newcommand{\ba}[1]{\begin{array}{#1}}
\newcommand{\ea}{\end{array}}

\newcommand{\bt}[1]{\begin{table}[ht]\centering\begin{tabular}{#1}}
\newcommand{\et}[1]{\end{tabular}\caption{\small#1}\end{table}}

\begin{document}

\title{Effect of Pions in Cosmic Rays}


\author{P. Castelo Ferreira}{
  address={CENTRA, Instituto Superior T\'ecnico, Av. Rovisco Pais, 1049-001 Lisboa, Portugal},
  email={pedro.castelo.ferreira@ist.utl.pt},
  thanks={SFRH/BPD/34566/2007}}

\author{J. Dias de Deus}{
  address={CENTRA, Instituto Superior T\'ecnico, Av. Rovisco Pais, 1049-001 Lisboa, Portugal},
  email={jdd@fisica.ist.utl.pt}}

\begin{abstract}
The effects of pions for vacuum polarization in background magnetic fields
are considered. The effects of quark condensates is also briefly addresses.
Although these effects are out of the measurement accuracy of laboratory
experiments they may be relevant for gamma-ray burst propagation. In particular,
for emissions from the center of the galaxy, we show that the mixing between
the neutral pion and photons results in a deviation of the gamma-ray spectrum from the
standard power-law in the TeV range.
\end{abstract}

\date{\today}

\maketitle

\section{Diagrams}

When background magnetic fields are present, traveling radiation interacts perturbatively
with those fields. The main two processes we are dealing in this presentation are
charged fermion/boson virtual loops~\cite{HE} and neutral scalar/pseudo-scalar exchange~\cite{pscalar}.
The respective diagrams and allowed intermediate particles are presented in table~\ref{dt}.

\begin{table}[h]
\begin{tabular}{ll}
\tablehead{1}{c}{b}{Virtual Loops}&\tablehead{1}{c}{b}{Scalar/Pseudo-scalar Exchange}\\
\includegraphics[width=60mm,viewport=130 370 510 700,clip]{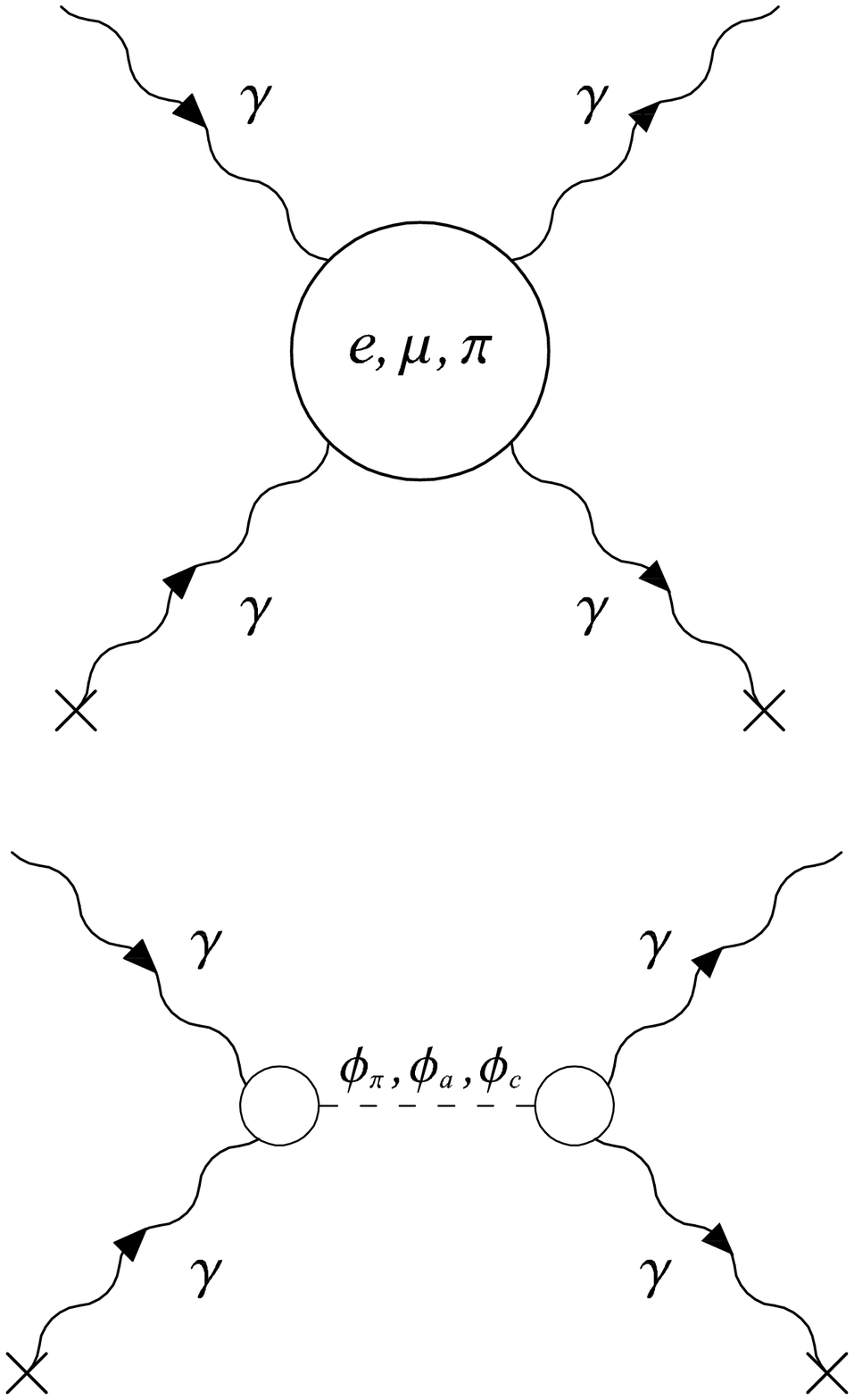}&\includegraphics[width=60mm,viewport=120 50 500 350,clip]{qq_diag.eps}\\
\ \ \ \ \ - Electron-Positron Loops ($e\bar{e}$)&\ \ \ \ \  - Neutral Pion (pseudo-scalar: $\phi_\pi$)\\
\ \ \ \ \ - Muon-Antimuon Loops ($\mu\bar{\mu}$) &\ \ \ \ \  - Axion (pseudo-scalar: $\phi_a$)\\
\ \ \ \ \ - Scalar Mesons Loops ($\pi^+\pi^-$)   &\ \ \ \ \  - Quark Condensates (scalar: $\phi_c$)\\
\ \ \ \ \ \ \ (plus quark condensates $\langle q\bar{q}\rangle$)
\end{tabular}
\caption{Diagrams and respective allowed processes for photon interaction with background electromagnetic fields.\label{dt}}
\end{table}

The processes for virtual loops are described by the Euler-Heisenberg Lagrangians, while
the processes for particle exchange are described by the interaction Lagrangians coupling
the scalar/pseudo-scalars to the gauge connection. We address these in the next sections following~\cite{paper}.

\section{Euler-Heisenberg Lagrangians}

\begin{figure}[h]
\includegraphics[width=60mm,viewport=30 267 430 550,clip]{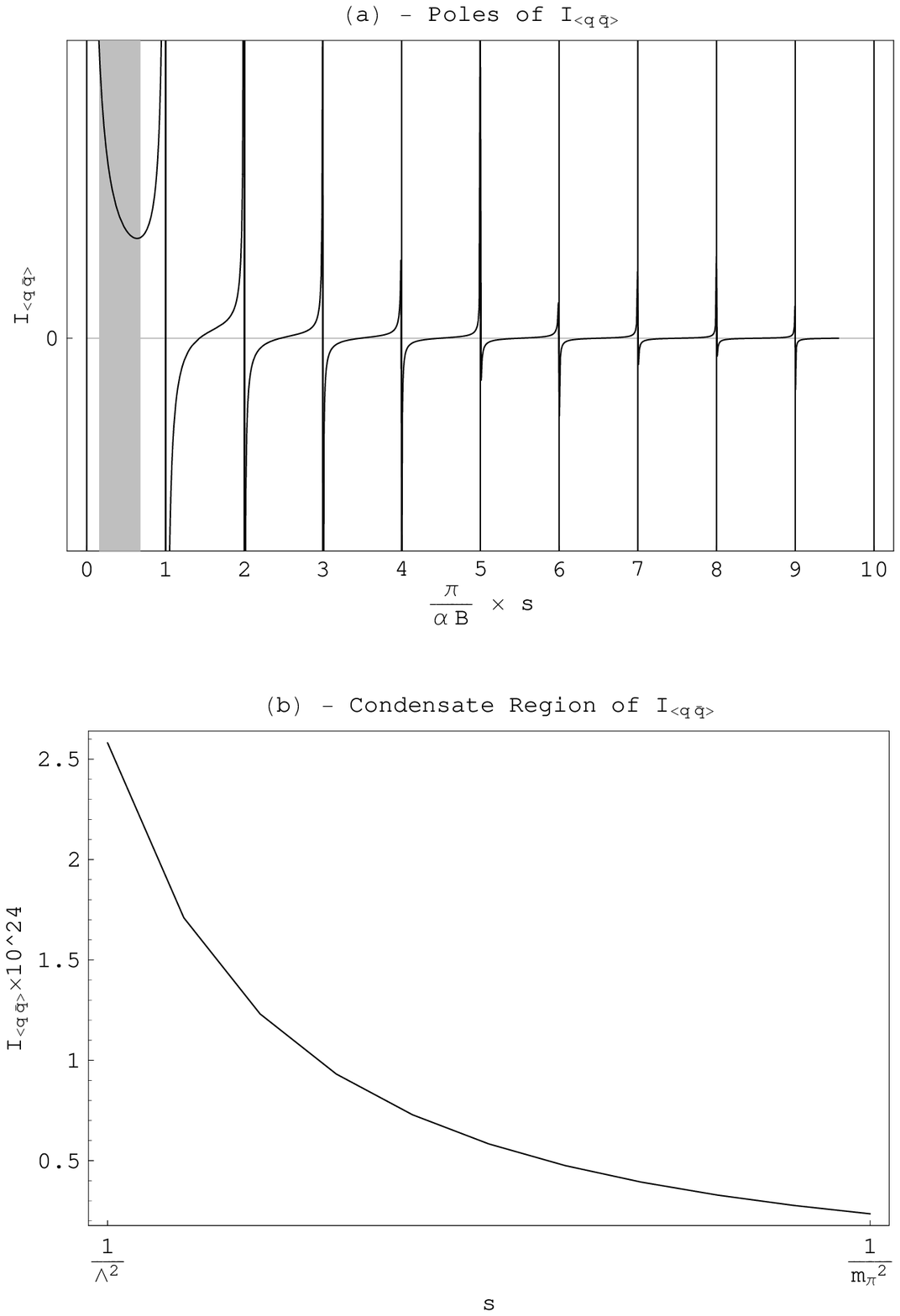}\ \ \includegraphics[width=60mm,viewport=35 -10 425 270,clip]{distribution.eps}
\caption{(a) The integrand~(\ref{int_qq}). The sum over the poles at $s=(n-1) \pi/\alpha B$ ($n>0$)
gives the pion vacuum polarization; (b) The same integrand for energies between $m_\pi=135\,MeV$ and $\Lambda=300\,MeV$ as
marked in (a) giving the quark condensate vacuum polarization contribution.\label{fig.poles}}
\end{figure}

For electron-positron virtual loops $e\bar{e}$ we have the Euler-Heisenberg Lagrangian~\cite{HE}
\be
\ba{rcl}
{\mathcal{L}}^{(2)}_{e\bar{e}}&=&\xi_e\left[4\left(F_{\mu\nu}F^{\mu\nu}\right)^2+7\left(\epsilon^{\mu\nu\delta\rho}F_{\mu\nu}F_{\delta\rho}\right)^2\right]\ ,\\[5mm]
\xi_e&=&\displaystyle\frac{2\alpha}{45\,(B^e_c)^2}=1.32\times 10^{-24}\,T^{-2}\ ,\ B^e_c=\frac{m_e^2 c^2}{e\,\hbar}\ .
\ea
\lb{L_e}
\ee
Where as usual $\alpha=1/137$ is the fine-structure constant, $e$ and $m_e=0.5\,MeV$ the charge and mass of the electron, $c$
the speed of light and $\hbar$ the Planck constant. The remaining contributions considered in this work will be
given relatively to this one. For muon virtual loops $\mu\bar{\mu}$ we have the Lagrangian
\be
\ba{rcl}
{\mathcal{L}}^{(2)}_{\mu\bar{\mu}}&=&\xi_\mu\left[4\left(F_{\mu\nu}F^{\mu\nu}\right)^2+7\left(\epsilon^{\mu\nu\delta\rho}F_{\mu\nu}F_{\delta\rho}\right)^2\right]\ ,\\[5mm]
\xi_\mu&=&\displaystyle\Delta_\mu\,\xi_e\ ,\ \Delta_\mu=\frac{\xi_\mu}{\xi_e}=\left(\frac{m_e}{m_\mu}\right)^4=5.43\times 10^{-10}\ ,
\ea
\lb{Pi2_mumu}
\ee
with $m_\mu=105\,MeV$ the muon mass. For charged pion loops $\pi^+\pi^-$ we have the Lagrangian
\be
\ba{rcl}
{\mathcal{L}}^{(2)}_{\pi^+\pi^-}&=&\xi_\pi\left[7\left(F_{\mu\nu}F^{\mu\nu}\right)^2+4\left(\epsilon^{\mu\nu\delta\rho}F_{\mu\nu}F_{\delta\rho}\right)^2\right]\ ,\\[5mm]
\xi_\pi&=&\displaystyle\Delta_\pi\,\xi_e\ ,\ \Delta_\pi=\frac{\xi_\pi}{\xi_e}=\frac{1}{2}\left(\frac{m_ef_\pi}{m^2_\pi}\right)^4=2.29\times 10^{-11}\ ,
\ea
\lb{Pi2_mumu}
\ee
where the pion constant is $f_\pi=93\,MeV$ and we take the pion mass to be $m_\pi=135\,MeV$.
There is one further contribution one can consider. The full distribution
for the pion loops is given by the integral
\be
\Pi_{\left\langle q\bar{q}\right\rangle}=\int_0^\infty ds\, I_{\left\langle q\bar{q}\right\rangle}\ ,\ I_{\left\langle q\bar{q}\right\rangle}=-\frac{\alpha\,B}{12\,f_\pi^4}\,\frac{1}{s^2}\left[\alpha\,B\,\cot(\alpha\,B\,s)-\frac{1}{s}\right]\ ,
\lb{int_qq}
\ee
represented in figure~\ref{fig.poles}. The sum over the poles gives the pion loop contribution, while the
the region between the first poles at $s=0$ and $s=\pi/\alpha B$ gives the quark condensate contribution
\be
\xi_c=\Delta_c\xi_e\ ,\ \Delta_c=\frac{\xi_c}{8\xi_e}=\frac{15\,m_e^4}{128\,f_\pi^4}\ln\left(\frac{\Lambda^2}{m_\pi^2}\right)\approx 1.69\times 10^{-10}\ ,
\ee
where we have taken the QCD cut-off to be $\Lambda\approx 300\,MeV$. This result was originally
computed within ChPT framework in~\cite{ChPT} and within NLJ framework in~\cite{NLJ}. In addition
we note that the quark condensates may only exist when very high densities of energy are
present $\left\langle E\right\rangle \sim 300 MeV/fm^3$~\cite{ChPT}. These values are only accessible
in very dense plasmas, for example in neutron stars~\cite{neutron} or near the center of galaxy~\cite{B_center}.

\section{Scalar/Pseudo-Scalar Exchange}

The Lagrangian for the pseudo-scalar neutral pion $\pi^0$ exchange is given by the
Adler-Bell-Jackiw anomaly~\cite{pscalar,ABJ}
\be
{\mathcal{L}}^{(2)}_{\pi^0}= \frac{1}{4}g_{\pi\gamma\gamma}\,\phi_{\pi^0}\,\epsilon^{\mu\nu\lambda\rho}F_{\mu\nu}F_{\lambda\rho}\ ,\ \
g_{\pi\gamma\gamma}=\frac{\alpha}{\pi f_\pi}=2.49\times 10^{-2}\,GeV^{-1}\ .
\lb{L_pi0}
\ee
We note that for the theoretical suggested axion the same Lagrangian applies
with the respective appropriate coupling constant $g_{a\gamma\gamma}$~\cite{pscalar}.
In addition we can take the quark condensate discussed in the previous section
as an effective scalar particle $\phi_c$ described by the following Lagrangian
\be
{\mathcal{L}}^{(2)}_c= \frac{1}{4}g_{c\gamma\gamma}\,\phi_c\,F_{\mu\nu}F^{\mu\nu}\ ,\ \ g_{c\gamma\gamma}=\Delta_c\xi_e\approx 1.69\times 10^{-10}\ .
\lb{L_pi0}
\ee
The other neutral meson contributions are lower by several orders of magnitude, although
their coupling constants to photons are of the same order of magnitude of the one for
$\pi^0$, their masses are higher~\cite{mesons}.

\section{Vacuum Birefringence}

\begin{figure}[h]
\includegraphics[width=60mm]{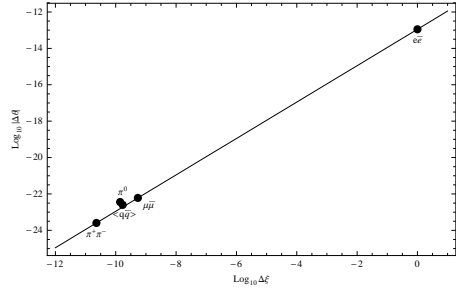}
\caption{Relative polarization rotation~(\ref{rot}) given in terms of the $\Delta\xi_i$ for
electron-positron loops ($e\bar{e}$), muon-antimuon loops ($\mu\bar{\mu}$) interchange of the neutral pion ($\pi^0$), quark condensates ($\left\langle q\bar{q}\right\rangle$) and charged pion loop ($\pi^+\pi^-$).\label{fig.effects_pol}}
\end{figure}

For radiation traveling in a background magnetic field $B_0$, due to the radiative
corrections discussed in the previous sections, it is induced a birefringent vacuum dispersion relation~\cite{disp}
\be
\ba{rcl}
\omega_{\perp,\parallel}&=&k\left(1-\lambda_{\perp,\parallel}B_0^2\right)\ ,\\[5mm]
\lambda_\perp&=&\displaystyle 8\,\left(1+\Delta\xi_\mu+\Delta\xi_c+\frac{7}{4}\Delta\xi_{\pi^\pm}\right)\,\xi_e\,B_0^2\ ,\\[5mm]
\lambda_\parallel&=&\displaystyle 14\,\left(\vphantom{\frac{4}{7}}1+\Delta\xi_\mu+\Delta\xi_{\pi^0}+\frac{4}{7}\Delta\xi_{\pi^\pm}\right)\,\xi_e\,B_0^2\ ,
\ea
\lb{Pi2_ee}
\ee
which induces both a polarization rotation $\Delta\theta$ and an ellipticity $\psi$
\be
\Delta\theta=\frac{1}{4}\left(\lambda_\parallel-\lambda_\perp\right)\Delta z\,\sin(2\theta_0)\ \ ,\ \ \ \psi=-\omega\,\Delta\theta.
\lb{rot}
\ee
Are represented in figure~\ref{fig.effects_pol} the relative magnitude of the polarization rotation
induced by the several radiative corrections discussed in this work. We note that today's laboratory
experiments accuracy is not sensitive to any of these corrections.

\section{$\gamma$-Ray Propagation}

\begin{figure}[h]
\includegraphics[width=60mm]{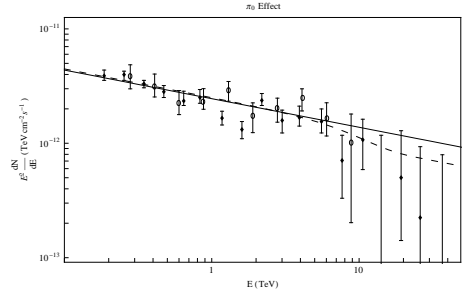}
\caption{Dashed line represents the deviation to the power law (continuous line)
due to the photon mixing to $\pi^0$. The data points are from the HESS col. (July/august 2003/2004)~\cite{HESS}.\label{fig.effects_sep}}
\end{figure}

The results of the previous sections can also be applied
to high energy $\gamma$-ray bursts. In order to do so
consider the propagation equations for photons in
background fields~\cite{Raffelt}
\be
\left(\omega-i\partial_z+M\right)\left[\ba{c}A_\parallel\\[3mm]A_\perp\\[3mm]\phi\ea\right]=0\ ,\ \
M=\left[\ba{ccc}
\Delta_{\gamma\gamma}+\Delta_\parallel&0&\Delta^\parallel_{\gamma\phi}\\[3mm]
0&\Delta_{\gamma\gamma}+\Delta_\perp&\Delta^\perp_{\gamma\phi}\\[3mm]
\Delta^\parallel_{\gamma\phi}&\Delta^\perp_{\gamma\phi}&\Delta_{\phi}
\ea\right]\ ,
\lb{matrix}
\ee
with the several entries given by
\begin{equation}
\Delta_{\gamma\gamma}\approx -i\frac{\Gamma}{2z_0}\ln\left(E\right)\ ,\ \Delta_\parallel\approx 4\xi_e B^2\ ,\ \Delta_\perp\approx 7\xi_e B^2\ ,\ \Delta^{\parallel,\perp}_{\gamma\phi}=\frac{1}{2}g_{\phi\gamma\gamma}B^{\parallel,\perp}\ ,\ \Delta_{\phi}=m_{\phi}\ .
\end{equation}
We will address radiation from the center of the galaxy at a distance of $z_0=8.5\,kpc$~\cite{gamma_conv,HESS}.
In the last expression $E$ stands for the $\gamma$-ray energy and we took the standard power
law approximation valid for energies in the $TeV$ range corresponding to $\Gamma=-\ln(dN/dE)\approx 2.25$~\cite{HESS}
for which the main contributions are due to photon desintegration~\cite{desintegration}.
For non-polarized radiation in gaussian magnetic field
distributions in domains of average size $s$, the conversion probability of photons to
pseudo-scalars is in the saturated continuum limit $z\gg s$
\be
P_{\gamma\to\phi}=\frac{1}{3}\left(1-e^{-\frac{3P_0z}{2s}}\right)\ ,\ \ P_0\approx 0.4\times 10^{-7}\left(\frac{g\,B_G E_{10}}{m_\phi^2}\right)^2\ .
\lb{prob}
\ee
The only measurable effect from the ones discussed in this work in the $TeV$ range is due to
the photon mixing to the neutral pion. Hence, following~\cite{Raffelt}, we are taking the
root mean square magnetic field strength $B_G=1\,\mu Gauss$, the radiation energy $E_{10}$
given in units of $10\,TeV$, for a distance $z=z_0=8.5\,kpc$, the domain size $s=0.01\,pc$ and $g=2.49\times 10^4$ with
$m_\phi=m_{\pi^0}=135\,MeV$. The resulting deviation to the power law is represented in figure~\ref{fig.effects_sep}.

\section{Conclusions}

We have shown that the exchange of the neutral pion $\pi^0$ with photons hold a deviation
from the power law spectrum for $\gamma$-ray from the center of the galaxy in the $TeV$.
This result may improve our knowledge of the effects affecting $\gamma$-ray bursts,
hence allowing a better understanding of its characteristics at the origin.

In addition the propagation equation~(\ref{matrix}) is also valid for light quark condensates ($m_c\approx 20\,MeV$).
Up to distances of $z\approx 125\,pc$ from the center of the galaxy the necessary energy densities for its existence
are present~\cite{B_center}. However the perturbative saturated limit is only applicable in the $GeV$
range considering, for example, a domain size of $s=.01 pc$~\cite{HESS2} its effects are observable. In the $TeV$
range the probability $P_0$ as given in~(\ref{prob}) does not preserve unitary, hence it is not applicable.
One can also add the effect of the axion~\cite{Raffelt} to the one from the $\pi^0$ which result should be to further
reduce the spectrum over energies of $E>10\,TeV$. Also it is expected that the effects discussed in this work are
relevant near neutron stars due to the high magnetic fields present in such environments~\cite{neutron}.\\

\noindent Acknowledgments -- Work of PCF supported by SFRH/BPD/34566/2007.

\end{document}